\documentclass[aps,prb,reprint,superscriptaddress,showpacs]{revtex4-1}
\usepackage{graphicx}			
\usepackage{bm}				
\usepackage{amsmath}			
\usepackage{amsthm}			
\usepackage{amssymb}			
\usepackage{multirow}

\begin{document}

\title{Electronic and optical properties of GaSb:N from first principles}

\author{Priyamvada Jadaun}
\email[]{priyamvada@utexas.edu}
\homepage[]{https://webspace.utexas.edu/~pj3292}
\affiliation{Cornell University, Ithaca, NY, 14850}

\author{Hari P. Nair}
\affiliation{Cornell University, Ithaca, NY, 14850}

\author{Vincenzo Lordi}
\affiliation{Lawrence Livermore National Laboratory, Livermore, California 94550, USA}

\author{Seth R. Bank}
\affiliation{Microelectronics Research Center, The University of Texas at Austin, Austin, TX 78758}

\author{Sanjay K. Banerjee}
\affiliation{Microelectronics Research Center, The University of Texas at Austin, Austin, TX 78758}

\date{\today}

\begin{abstract}
GaSb:N displays promise towards realization of optoelectronic devices accessing the mid-infrared wavelength regime. Theoretical and experimental results on its electronic and optical properties are however few. To address this, we present a first principles, density functional theory study using the hybrid HSE06 exchange-correlation functional of GaSb doped with 1.6$\%$ nitrogen. To study dilute-nitrides with small band gaps, the local density approximation (LDA) is insufficient and more accurate techniques such as HSE06 are needed. We conduct a comparative study on GaAs:N, also with 1.6$\%$ nitrogen mole fraction, and find that GaSb:N has a smaller band gap and displays more band gap bowing than GaAs:N. In addition we examine the orbital character of the bands, finding the lowest conduction band to be quasi-delocalized, with a large N-$3s$ contribution. At high concentrations, the N atoms interact via the host matrix, forming a dispersive band of their own which governs optoelectronic properties and dominates band gap bowing. While this band drives the optical and electronic properties of GaSb:N, its physics is not captured by traditional models for dilute-nitrides. We thus propose that a complete theory of dilute-nitrides should incorporate orbital character examination, especially at high N concentrations. 
\end{abstract}


\maketitle

\section{Introduction}

	Dilute-nitride III-V alloys provide unique opportunities for band gap engineering\cite{veal_band_2005, belabbes_giant_2006, harris_development_2007}. In these materials, substituting a small percentage of the column-V element with nitrogen leads to anomalous band gap bowing\cite{shan_band_1999, kondow_gainnas:_1996}. It is well known that N displays this interesting property due to its small size and electronegativity mismatch with the host matrix, thereby creating a localized but powerful disturbance in the host electronic potential\cite{gueddim_alloy_2007, shan_band_2004}. While the general theory of anomalous band gap reduction in GaAs-based dilute-nitrides (GaAs:N) has been well studied, many effects and concentration regimes remain unexplored. 

	Moreover, GaSb:N, also a dilute nitride III-V, displays a large anomalous reduction in the band gap, due to the considerable electronegativity mismatch between N and Sb\cite{belabbes_giant_2006}. This makes GaSb:N a promising material for optoelectronics accessing the 2-5 $\mu$m mid-infrared wavelength regime\cite{lindsay_theory_2008, veal_band_2005, belabbes_giant_2006, nair_structural_2012}. However, despite the keen interest, reports on GaSb:N optical experiments are few and theoretical studies even fewer\cite{gueddim_alloy_2007, jefferson_band_2006, veal_band_2005, belabbes_giant_2006}. The reported radiative efficiency of GaSb:N is relatively weak\cite{wang_band_2009, iyer_effects_2007}, leading to uncertainty regarding its suitability for optoelectronic applications. In order to understand and predict its optical properties, a reliable and detailed theoretical model is essential. 

	In this paper, we describe an \textit{ab initio} study of the electronic and optical properties of GaSb:N using density functional theory (DFT) with the accurate Heyd-Scuseria-Ernzerhof (HSE) hybrid exchange-correlation functional. We also compare the optoelectronic properties of GaSb:N with GaAs:N, the latter being well-studied and find good agreement with published GaAs:N results\cite{lordi_nearest-neighbor_2003}. Moreover, we present an accurate structural analysis and a detailed study of the orbital character of the bands. 
	
	We find that at concentrations as high as 1.6$\%$, the N atoms interact with one another via the host matrix, forming a quasi-delocalized band, an effect not captured by traditional band gap bowing models.  This is a vital omission since this band provides the main mechanism for band gap reduction and dominates the optoelectronic properties of GaSb:N. We see similar behaviour in GaAs:N, reported earlier\cite{lordi_nearest-neighbor_2003}, which points to the ubiquitous nature of this phenomenon. We thus propose that the theory for dilute-nitrides should be extended to incorporate orbital character and N-N interactions, especially at N concentrations ranging from $1-2\%$\cite{lordi_nearest-neighbor_2003}. Upon comparison, we notice that the optical matrix element for GaSb:N is weaker than for GaAs:N. However, with the removal of N related clusters and defects which lead to non-radiative recombination pathways\cite{wang_band_2009, kudrawiec_conduction_2011, lindsay_theory_2008}, GaSb:N could be a promising material for optoelectronic applications. 

\begin{figure}[t]
\includegraphics[width=\columnwidth]{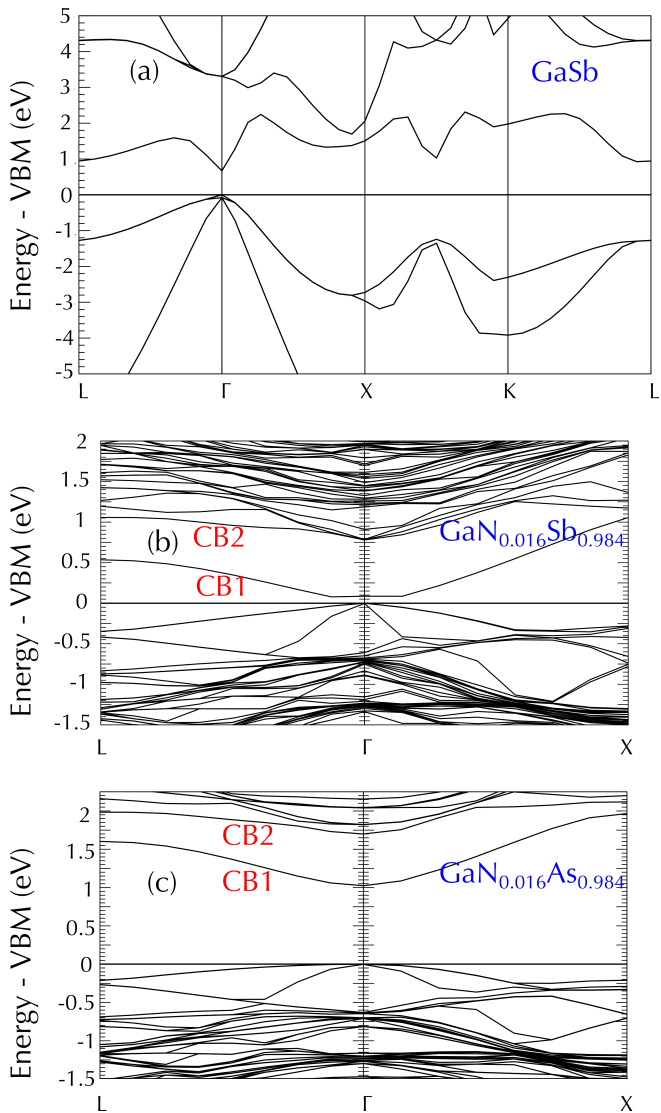}
\caption{(Color online) Band structure calculated with HSE06 for (a) bulk GaSb (b) GaSb:N doped 1.6$\%$ with N and (c) GaAs:N doped 1.6$\%$ with N. 
The zero of energy is set to the valence band maximum (VBM) in each case. The important valence and conduction bands are labelled.\label{fig1}}
\end{figure}

\section{Calculations} 
	
	\textit{Ab initio} calculations were performed with the Vienna-\textit{Ab Initio} Simulation Package (VASP)\cite{VASP1, VASP2, VASP3} using the projector augmented wave (PAW) method\cite{PAW} and pseudopotentials and employing the range-separated hybrid functional of Heyd, Scuseria, and Ernzerhof (HSE06)\cite{HSE3}. The atomic relaxations for GaSb:N were performed using the Perdew-Burke-Ernzerhof (PBE) parametrization of the generalized gradient approximation (GGA)\cite{GGA-PBE1, GGA-PBE2}. Bulk GaSb has a 2 atom primitive unit cell, while GaN$_{0.016}$Sb$_{0.984}$ requires a 128 atom supercell. Calculations were also extended to GaN$_{0.008}$Sb$_{0.992}$ (256 atoms). Starting with the experimental bulk parameters for GaSb, the supercell structures were allowed to relax until the Hellman-Feynman forces on all atoms were less than 0.01 eV/\AA; the cell volume was optimized via fitting to the Birch-Murnaghan equation of state. Full relaxation using HSE06  is computationally prohibitive and was only performed on bulk GaSb. Relaxed structures for GaSb:N were obtained by using PBE to relax the atoms and using HSE06 bulk lattice parameters to estimate the relaxed volumes by extrapolation. Brillouin zone sampling for the initial LDA calculations used 4$\times$4$\times$4 \textbf{k}-mesh followed by 3$\times$2$\times$3 for the PBE calculations and 1$\times$1$\times$1 or 2$\times$2$\times$2 for HSE06. The energy cut off for the plane wave basis was 500 eV for hybrid calculations. Convergence was checked with respect to higher \textbf{k}-mesh and energy cut off values. GaN$_{0.016}$As$_{0.984}$, being just a reference, was relaxed using the local density approximation (LDA)\cite{LDA} for the exchange-correlation functional. 

	After optimization, bulk GaSb was found to display a lattice constant of 6.16 \AA\ which is comparable to the experimental value of 6.096 \AA\cite{ioffe}. Relaxation of N and the surrounding atoms in GaN$_{0.016}$Sb$_{0.984}$, GaN$_{0.008}$Sb$_{0.992}$, and GaN$_{0.016}$As$_{0.984}$ all followed the same pattern. The 4 nearest neighboring Ga atoms were pulled closer to N, to adjust to the small bond length of Ga--N. To compensate, the bond lengths for the next-neighboring shell of atoms (Sb or As) were larger than average and they approached their bulk (GaSb or GaAs) values farther away from the N atom. The Ga--N bond length in GaN$_{0.016}$Sb$_{0.984}$ was 2.09 \AA\ and in GaN$_{0.008}$Sb$_{0.992}$ was 2.11 \AA. These values are close to the bulk Ga--N bond length of 1.96 \AA\cite{ioffe} and much smaller than the bulk Ga--Sb bond length of 2.66 \AA. Similarly in GaN$_{0.016}$As$_{0.984}$, the relaxed Ga-N bond length was 2.07 \AA\ while the bulk Ga-As bond length is 2.45 \AA\cite{ioffe}. This gives us an idea of the large strain in the system introduced by the presence of a N atom. 

\begin{figure}[t]
\includegraphics[width=\columnwidth]{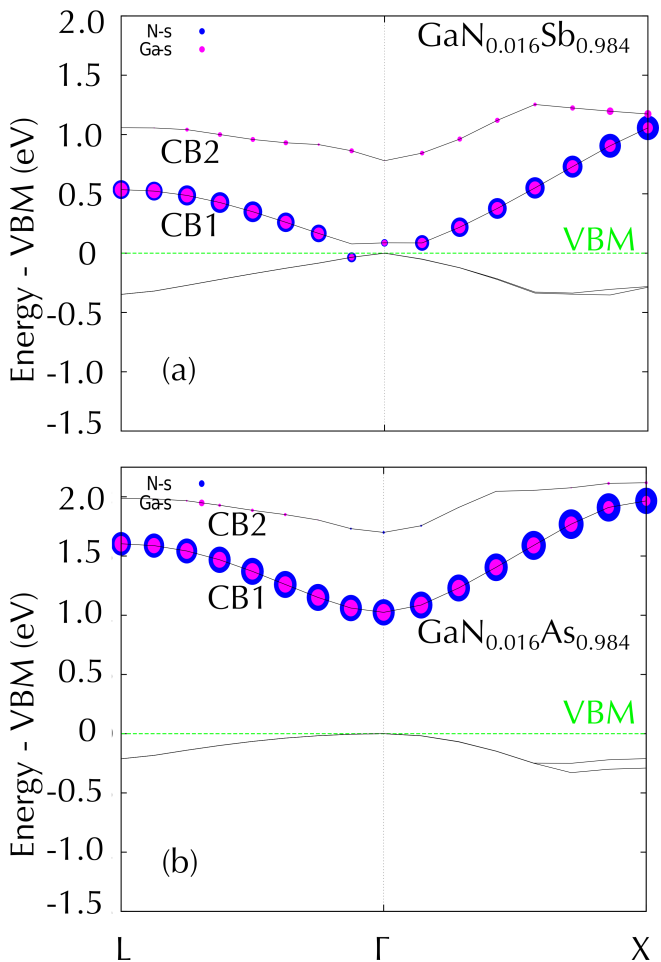}
\caption{(Color online) Orbital projected band structure for (a) GaN$_{0.016}$Sb$_{0.984}$ and (b) GaN$_{0.016}$As$_{0.984}$, projected on the N-$s$ orbital as well as the Ga-$s$ orbital for gallium atoms bonding to nitrogen. The projections are calculated per atom. There are 4 bands plotted i.e. VB2, VB1, CB1 and CB2 with VB1 and VB2 overlapping for the most part. Zero energy is at the valence band maxima (VBM).\label{fig2}}
\end{figure}

\section{Band gaps and orbital character} 
	
	The bulk GaSb band structure using HSE06 functionals, shown in Figure \ref{fig1}(a), displays a band gap of 749 meV, which is remarkably close to the experimental band gap of $\sim$ 720 meV\cite{veal_band_2005}. This is a testament to the accuracy of the HSE06 functional for these materials, given the well-known problem of the underestimation of band gaps by LDA/GGA. Band structure for GaN$_{0.016}$Sb$_{0.984}$ was also obtained (Figure \ref{fig1}(b)) and displays a direct gap at $\Gamma$ of 86.6 meV (reduction of 662 meV from bulk GaSb) with the L valley lower than the X, unlike a previous report which suggested an indirect gap\cite{gueddim_alloy_2007}. At concentrations of $0.8\%$ this gap increases to 192 meV. Experimental data for GaSb:N shows a band gap of $\sim$400 meV for $1.6\%$\cite{wang_band_2009, veal_band_2005} and that of $\sim$500 meV for $0.8\%$ concentration. While the agreement with experiments is less stark in the latter case, it is indeed better than the negative gaps produced by LDA. For GaN$_{0.016}$As$_{0.984}$ the band structure is also shown (Figure \ref{fig1}(c)) and displays a band gap of 1.024 eV (a reduction of $\sim$400 meV from bulk GaAs). The calculated band gap is close to the experimental value of $\sim 1.20$ eV\cite{bi_bowing_1997} and larger than that seen for GaSb:N, as is expected. Moreover, the band gap bowing in GaSb:N is greater than that seen in GaAs:N, which is also expected. This is due to the much larger electronegativity difference between N and Sb (in GaSb:N) as compared to that between N and As (in GaAs:N) atoms\cite{veal_band_2005}. 

\begin{figure}[t]
\includegraphics[width=\columnwidth]{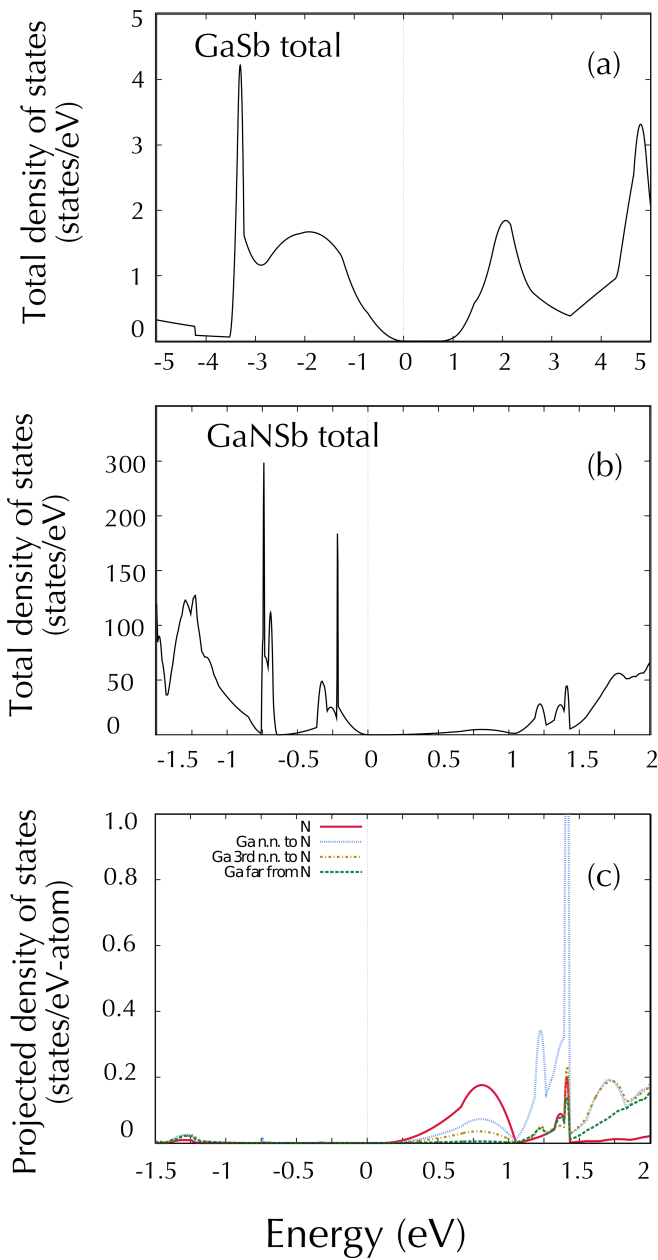}
\caption{(Color online) Total density of states plotted for (a) bulk GaSb and (b) GaN$_{0.016}$Sb$_{0.984}$ and (c) per atom density of states for the $s$ orbital in GaN$_{0.016}$Sb$_{0.984}$ projected on the N atom, a Ga atom bonding to N, a Ga atom 3 bonds away from N and a Ga atom far away from N.\label{fig3}}
\end{figure}

	We make particular note of the nature of the conduction bands in these dilute N materials. The lowest conduction band, which we call CB1, has a wide dispersion, i.e., close to 500 meV. Moreover, this dispersion is qualitatively similar to the higher lying conduction bands (i.e. it is not a flat band as one would expect from isolated N levels). This points towards the quasi-delocalized nature of CB1. To study the orbital character of these bands we plot the projected band structure (Figure \ref{fig2}) and the projected density of states(DOS) (Figure \ref{fig3}). Figure \ref{fig2}(a) plots the band structure for GaN$_{0.016}$Sb$_{0.984}$ as projected onto the N-$s$ orbital, as well as, the $s$ orbitals belonging to the gallium atoms that bond with nitrogen (i.e. nearest neighbors). The size of the symbols in the plot is proportional to the amount of orbital contribution to the band, at various \textbf{k}-points. For GaSb:N, amongst CB1 (lowest conduction band), CB2 (second lowest conduction band), VB1 (highest valence band) and VB2 (second highest valence band), it is CB1 which shows contributions from the $s$ orbital of the N atom. The per atom contributions coming from the $s$ orbitals of the Ga atoms that bond to N are also higher for CB1, although there is some contribution from the Ga-$s$ to CB2 as well (unlike N-$s$). The projected DOS (Figure \ref{fig3}(c)) shows clear and strong hybridization between the N-$s$ and neighboring Ga-$s$ signifying strong covalent bonding. This is corroborated by the structural evidence, we presented earlier, of Ga-N bond lengths in the alloys approaching the bulk value. Moreover, it is clear from Figure \ref{fig3}(c) that while the Ga-$s$ contribution from the atoms neighboring N is high in CB1, as we move away from the N atom in real space, the Ga-s contribution to CB1 drops and shifts to higher lying conduction bands. Similar behavior is seen for GaAs:N by us, and was also reported in an earlier study\cite{lordi_nearest-neighbor_2003}. 

	The projected band structure and DOS show that CB1 is a band that forms below the bulk conduction band edge (CBE), by $s$-orbital interactions between N and its nearest neighboring Ga atoms. However, the quasi-delocalized nature of the band cannot be explained by local N-Ga nearest neighbor interactions alone, which would give a very flat band dispersion. We thus conclude that the band CB1 also originates in long range interactions between the N atoms themselves mediated via the host matrix. The concentration of the N atoms at 1.6$\%$ is high enough for the propagation of such effects. This is similar to the picture presented by Lordi et al.\cite{lordi_nearest-neighbor_2003} to explain the spectrum of GaAs:N. Since CB1 is the lowest lying band, it therefore dominates band gap bowing and optoelectronic properties of GaSb:N, making accurate determination of its nature very important. 

	 To investigate optical properties further, we calculated the optical transition matrix elements for GaSb:N and GaAs:N using PBE. The results suggest that GaSb:N may have poorer optical absorption and emission compared to GaAs:N. It is also known that N related clusters and defects in GaSb:N form levels close to and even below the CBE\cite{wang_band_2009, kudrawiec_conduction_2011, lindsay_theory_2008} adversely impacting its luminescence efficiency.

	Traditionally, the anomalous band gap bowing behaviour of dilute-nitrides has been explained with a local band anti-crossing model(BAC)\cite{shan_band_2004, shan_band_1999}. For our structures, the BAC model would predict that, upon alloying with N, the lowest conduction band in GaSb splits into a lower lying band which is mainly host-like (or Ga-like) in character, with an upper band which is mainly N-like. It would also predict this upper, N-like, conduction band to be localized and non-dispersive. However, we observe most N-like character in the lowest conduction band which is also quasi-delocalized. The erronous prediction of BAC is understandable since at high concentrations of N, such as $1.6\%$, we are out of the ultra-dilute regime where the BAC model is accurate\cite{shan_band_2004}. This implies that, for the theoretical models to accurately capture the band structure properties, it is important to incorporate long-range N-N interactions. This is especially vital, since it is the lowest band (i.e. CB1) that determines the optoelectronic properties of dilute-nitrides. We thus propose that a full theory of dilute-nitrides should include an examination of the orbital character of the bands, specifically N-N long range interactions. 
	
\section{Summary} 
	
	In summary, we have presented a detailed hybrid functional-based DFT study of the electronic structure of GaSb:N. We found that GaSb:N behaves qualitatively similar to GaAs:N at 1.6$\%$ N concentration, but with slightly stronger band gap bowing. At such high concentrations, the N atoms interact with each another via the host matrix atoms. This leads to the formation of a quasi-delocalized, N band, which is also the lowest conduction band. This N band thus dictates the band edge optical properties of these dilute-nitrides and drives band gap bowing. Owing to the importance of this band, we emphasize the need to extend traditional dilute-nitride models to incorporate orbital character. While the optical matrix element of GaSb:N is not as high as that of GaAs:N, the removal of N-related defects could make it a promising material for optoelectronic devices.

\begin{acknowledgments}
We would like to acknowledge Texas Advanced Computing Center (TACC) for access to high performance computing resources.
Part of this work was performed under the auspices of the U.S. Department of Energy by Lawrence Livermore National Laboratory under Contract DE-AC52-07NA27344.
\end{acknowledgments}

\bibliography{GaNSb_tot}

\end{document}